\documentclass[twocolumn]{aastex62} 
\usepackage{graphicx}
\usepackage{txfonts}
\usepackage{threeparttable}

\received{xxx}
\revised{xxx}
\accepted{}
\submitjournal{ApJL}
\shorttitle{Thermal reprocessing and optical/UV lags in NGC~5548}
\shortauthors{Kammoun et al.}

\def\msun{\hbox{M$_\odot$}}

\def\cm3{\hbox{cm$^{-3}$}}

\def\mdot{\hbox{$\dot{m}$}}
\def\meantau{\hbox{$\langle \tau \rangle$}}
\begin{document} 

\title{A hard look at thermal reverberation and optical/UV lags in NGC~5548}
%\AuthorCallLimit = 
\correspondingauthor{E. Kammoun}
\email{ekammoun@umich.edu}

\author[0000-0002-0273-218X]{E. S. Kammoun}
\affiliation{Department of Astronomy, University of Michigan, 1085 South University Avenue, Ann Arbor, MI 48109-1107, USA}

\author{I. E. Papadakis}
\affiliation{Department of Physics and Institute of Theoretical and Computational Physics, University of Crete, 71003 Heraklion, Greece}
\affiliation{Foundation for Research and Technology - Hellas, IESL, Voutes, 71110 Heraklion, Greece}
\affiliation{Institute of Astrophysics, FORTH, GR-71110 Heraklion, Greece}

\author{M. Dov\v{c}iak}
\affiliation{Astronomical Institute of the Academy of Sciences, Bo{\v c}n{\'i} II 1401, CZ-14100 Prague, Czech Republic}

\begin{abstract}

The UV/optical variations in many AGN are very well correlated, showing delays which increase with increasing wavelength. It is thought that this is due to thermal reprocessing of the X-ray emission by the accretion disk. In this scenario, the variable X-ray flux from the corona illuminates the accretion disk where it is partially reflected, and partially absorbed and thermalized in the disk producing a UV/optical reverberation signal. This will lead to a time lag increasing with wavelength. However, although the shape of the observed time-lags as a function of wavelength is consistent with the model predictions, their amplitude suggested a disk which is significantly hotter than expected. In this work, we estimate the response functions and the corresponding time lags assuming a standard Novikov-Thorne accretion disk illuminated by a point-like X-ray source. We take into account all relativistic effects in the light propagation from the X-ray source to the disk then to the observer. We also compute the disk reflection, accounting for its ionization profile. Our results show that thermal reverberation effects are stronger in sources with large X-ray source height and low accretion rate. We also found that the time lags increase with height and accretion rate. We apply our model to NGC~5548 and we show that the observed lags in this source can be explained by the model, for a source height of $\sim 60 ~\rm r_g$ and an accretion rate of a few percent of the Eddington limit for a maximally-spinning black hole.

\end{abstract}


%
%-------------------------------------------------------------------

\section{Introduction}
\label{sec:intro}

Active galactic nuclei (AGN) are thought to be powered by accretion of matter onto a supermassive black hole (BH, with a mass $M_{\rm BH} \sim 10^{6-9}~\rm M_\odot$) in a form of an optically-thick, geometrically-thin disk \citep{Shakura73, Novikov73}. The disk emits a multi-temperature blackbody (BB) spectrum peaking in the optical/ultraviolet (UV) range. The temperature of the disk decreases with radius as $T(r) \propto r^{-3/4}$. A fraction of the disk photons are then Compton upscattered by a medium of hot electrons in the vicinity of the BH, the so-called `X-ray corona' \citep[e.g.,][]{Shapiro76, Haardt93}. Several lines of evidence are suggestive of a compact corona located at a few gravitational radii ($r_{\rm g} = GM_{\rm BH}/c^2$) above the BH \citep[e.g.,][]{Chartas09, Fabian2009, DeMarco2013, Reis13, Emmanoulopoulos2014}.

In this case, X-rays from the corona irradiate the accretion disk. Some of them will be reprocessed and re-emitted in the form of the disk `X-ray reflection spectrum', while the rest will be absorbed, and will increase the disk's temperature. As a result, the UV/optical emission of the disk will be enhanced. Most of the UV photons are expected to emerge from the hot inner regions, while the optical photons are expected to be emitted by the cool outer regions. Consequently, if the X-rays are variable, we expect the disk UV/optical emission to also vary with a time lag increasing with wavelength.

Several multi-wavelength monitoring campaigns, using the \textit{Neil Gehrels Swift Observatory} (hereafter \textit{Swift}), have been performed recently to study AGN variability across X-rays, UV and optical at high cadence and over long periods \citep[e.g.,][]{Mchardy14, Shappee14, Mchardy18, Cackett18, Edelson19}. In particular, \cite{Fausnaugh16} studied the X-ray/UV/optical lags in NGC~5548 using data from the \textit{Hubble Space Telescope (HST), Swift} and ground-based telescopes. The authors showed that the measured lags are in agreement with the predicted $\tau \propto \lambda^{4/3}$ relation, in the case of a standard Shakura-Sunyaev disk. However, they also found that the time lags were larger than expected, at all wavelengths. 

In this letter, we study the time-lags vs wavelength relation (hereafter ``lag-spectrum'') in the context of the lam-post geometry. We investigate the effects of the X-ray source height and the accretion rate on the reverberation signal and the lag spectra. Applying the model to NGC 5548, we find that a standard disk, with a low accretion rate ($\sim 0.005-0.01$ of the Eddington limit) is in agreement with the observed UV/optical lags in this source, as long as the X-ray source height is larger than $40~\rm r_g$.

%%%%%%%%% SECTION 2  
\section{Model setup}  
\label{sec:model}

We consider a Keplerian, geometrically-thin and optically-thick accretion disk, around a BH of mass $M_{\rm BH} $ and accretion rate \mdot. The disk is co-rotating with the BH and its temperature profile follows the Novikov-Thorne prescription \citep{Novikov73}, with a color temperature correction factor of 2.4. The disk extends from the innermost stable circular orbit (ISCO) at radius $r_{\rm ISCO}$, up to an outer radius of $r_{\rm out}=10^4 \rm r_g$. The ISCO radius is uniquely defined by the BH spin. In this work, we consider two extreme cases, one with $a^\ast = 0$ ($r_{\rm ISCO}= 6~\rm r_g$) and the other with $a^\ast = 1$ ($r_{\rm ISCO}= 1 ~\rm r_g$). We also assume a point-like X-ray source located at a height ($h$) on the rotational axis of the BH (i.e., the lamp-post geometry). The X-rays are emitted isotropically (in the rest frame of the lamp-post) with an intrinsic spectrum $F_{\rm X}(t) = N(t)E^{-\Gamma}\exp(-E/E_{\rm C})$, and illuminates the disk. Part of this flux is reprocessed and re-emitted in X-rays (this is the `disk reflection component') and part of it is absorbed. We assume that the X-ray source is variable in normalization only.

Let us assume that the X-ray source emits a flash with flux $F_{\rm X0}$ at time $t_0$, with a duration $\Delta t$. Hence, an incident primary flux $F_{\rm inc}(r,\tau')$ (in the disk's rest frame) will reach the disk at a radius $r$, and at a time $\tau'$. If $F_{\rm ref}(r,\tau')$ is the flux that is reflected from the disk, then 
\begin{equation}\label{eq1}
F_{\rm abs}(r,\tau') = F_{\rm inc}(r,\tau')-F_{\rm ref}(r,\tau'), 
\end{equation}
is the flux absorbed by the disk. This flux is then added to the original disk flux assuming a Novikov-Thorne profile, $F_{\rm NT}(r)$, and the sum can be used to estimate the disk temperature, as a function of radius and time, as follows,
\begin{equation}\label{eq2}
T_{\rm new}(r,\tau') = \left[ \frac{F_{\rm abs}(r,\tau')  + F_{\rm NT}(r) }{\sigma} \right]^{1/4},
\end{equation}
\noindent where $\sigma$ is the Stefan-Boltzmann constant.

%--------Fig 1
\begin{figure*}
\centering
\includegraphics[width = 0.48\linewidth]{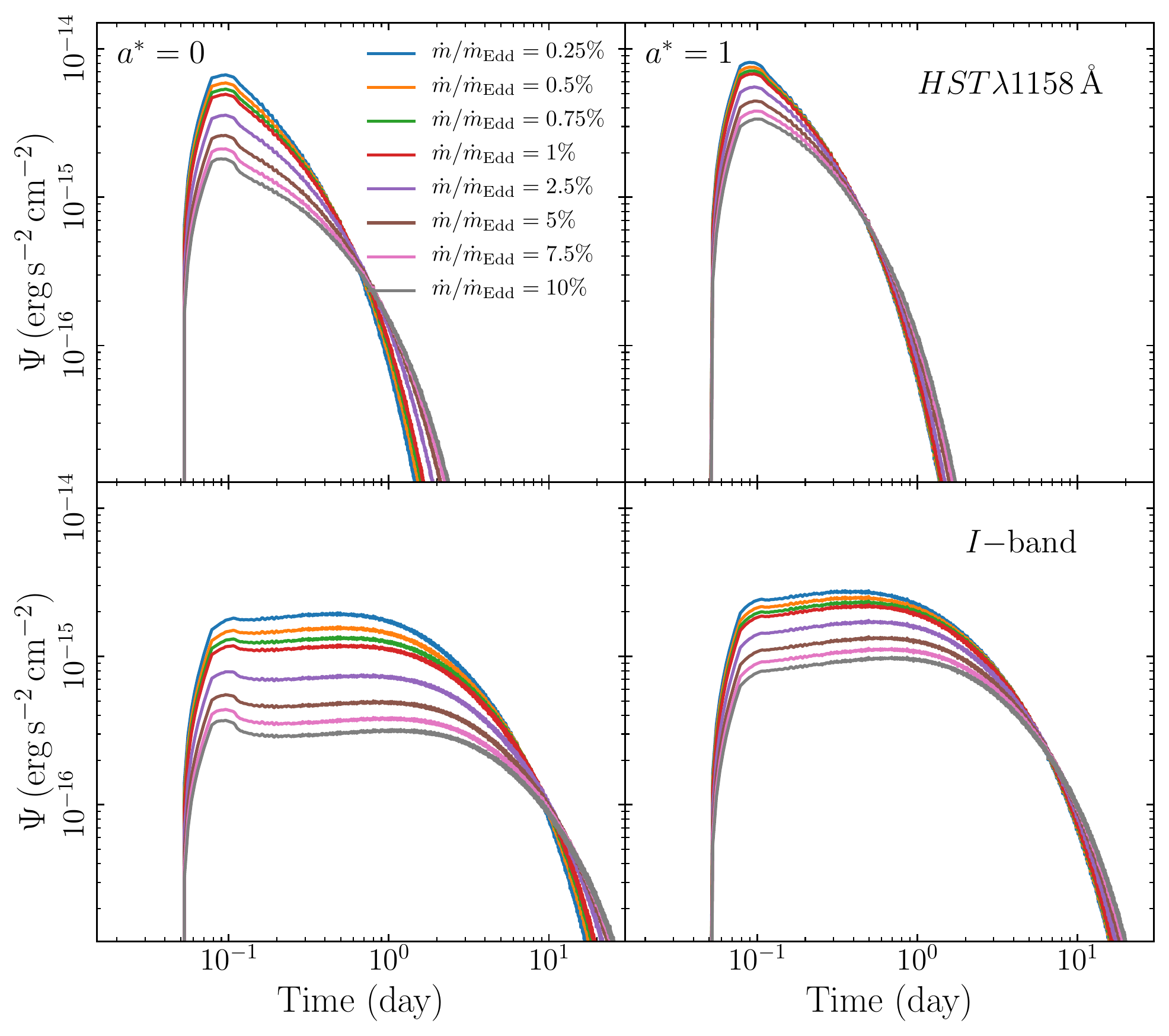}
\includegraphics[width = 0.48\linewidth]{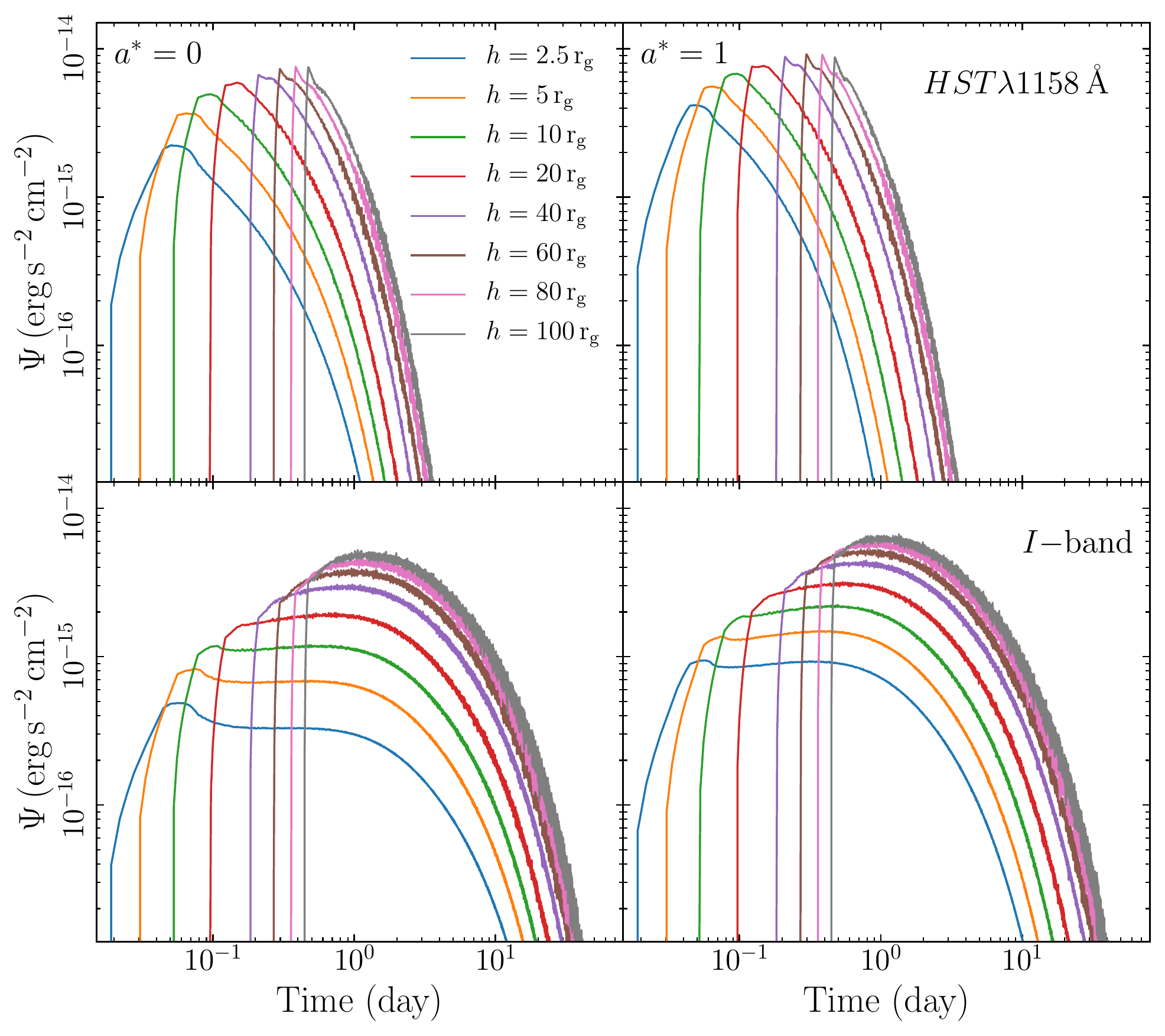}

\caption{Left panel: Response functions for the various \mdot\ values we considered ($h=10~\rm r_g$). Right panel: the same for the various $h$ values we considered ($\dot{m} =0.01~\dot{m}_{\rm Edd}$). Left and right columns correspond to $a^\ast =0$ and 1, respectively. Top and bottom rows show the responses for the {\it HST} $\lambda 1158$ and the $I$-band, respectively.}
\label{fig:responses}
\end{figure*}
%--------

Then we compute the disk response, $\Psi$, to the short X-ray flash. To do that we identify all the disk elements (in radius, $r$, and azimuth, $\varphi$) that a distant observer will ``see" to be illuminated at the same time. Note that the temperature of each of these elements, $T_{\rm new}$, will be different for each of them. Then we compute the flux that the observer receives, at a time $\tau_{\rm obs}$, from all these elements in a given waveband, $\Delta \lambda$, between, say, $\lambda_{\rm min}$ and $\lambda_{\rm max}$. Let us denote this flux as $F_{\rm rev}(\Delta\lambda, \tau_{\rm obs}$). Let us also denote with $F_{\rm NT}(\Delta\lambda)$, the flux of these elements, in the same band pass, when their temperature is equal to the Novikov-Thorne. We define the disk's ``response function" in a given waveband as follows,

\begin{equation}
\Psi(\Delta\lambda, \tau_{\rm obs}) = \frac{F_{\rm rev}(\Delta\lambda, \tau_{\rm obs}) - F_{\rm NT}(\Delta\lambda)}{F_{\rm X0}\,\Delta t},
\end{equation}
\noindent
for each time, $\tau_{\rm obs}$, as the illumination progresses across the disk. The equation above shows the ``extra" flux that is emitted by the disk, at each time $\tau_{\rm obs}$ (in the observer's frame), due to the heating caused by the absorption of the incident X-rays. We note that the response function is normalized to the observed X-ray flux. The total (observed) flux emitted by the disk in the $\Delta\lambda$ band, and at time $\tau_{\rm obs}$ will then be equal to,

\begin{equation}
F_{\rm obs}(\Delta\lambda, \tau_{\rm obs}) = F_{\rm NT}(\Delta \lambda) +
\int_{-\infty}^{\infty}F_{\rm X}(t')\Psi[\Delta\lambda, (\tau_{\rm obs}-t')]{\rm d}t'.
\end{equation}

All the computations mentioned above were performed using the {\tt KYNXILREV}\footnote{\url{https://projects.asu.cas.cz/stronggravity/kynreverb/}} model (M. Dov\v{c}iak et al., in prep.). Given the {\it observed} 2--10 keV band luminosity of the X--ray source, the model estimates the intrinsic luminosity and the incident flux on each disk radius, as a function
of time (in the observer's frame), taking into account all the relativistic effects in the propagation of light from the primary source to the disk. Given the incident X-ray flux on the disk at each radius, the model estimates the radial ionization profile of the accretion disk assuming a constant disk density\footnote{The choice of a constant disk density in the estimation of the ionization disk profile should not significantly affect our results because the radial dependence of any realistic density profile is much less significant than the radial decrease of the disk illumination by the lamp-post \citep[see e.g.][]{Svoboda12, Kammoun19}.} \citep[see][for more details about the ionization estimates]{Kammoun19}. The disk reflection spectrum is computed using the {\tt XILLVERD} tables for the reflection spectrum from ionized material \citep{Garcia16}, by integrating them from 0.1~keV to infinity. The code then estimates $F_{\rm abs}(r,\tau')$, $T_{\rm new}(r,\tau')$ and, finally $\Psi(\lambda, \tau)$, taking into account all relativistic effects in the propagation of light from the X-ray source to the disk, and from the disk to the observer.

%%%%%%%%% SECTION 3
\section{The disk response}
\label{sec:Simulations}

To compute the disk response we chose model parameters values that are applicable for NGC 5548. In particular, we assumed an $ M_{\rm BH} = 5 \times 10^7$~\msun\ \citep{Bentz15}, and an inclination of 40\degr. Using the results presented by \cite{Mathur17}, assuming a power-law photon index $\Gamma = 1.5$, we estimate the observed 2-10~keV luminosity of the source to be $L_{\rm X}/L_{\rm Edd} = 0.0034$. We also assumed a high-energy cutoff of 300~keV, and a luminosity distance of 75~Mpc as listed in the {\tt Simbad} database \citep{Simbad}.

%--------Fig 2
\begin{figure*}
\centering
\includegraphics[width = 0.49\linewidth]{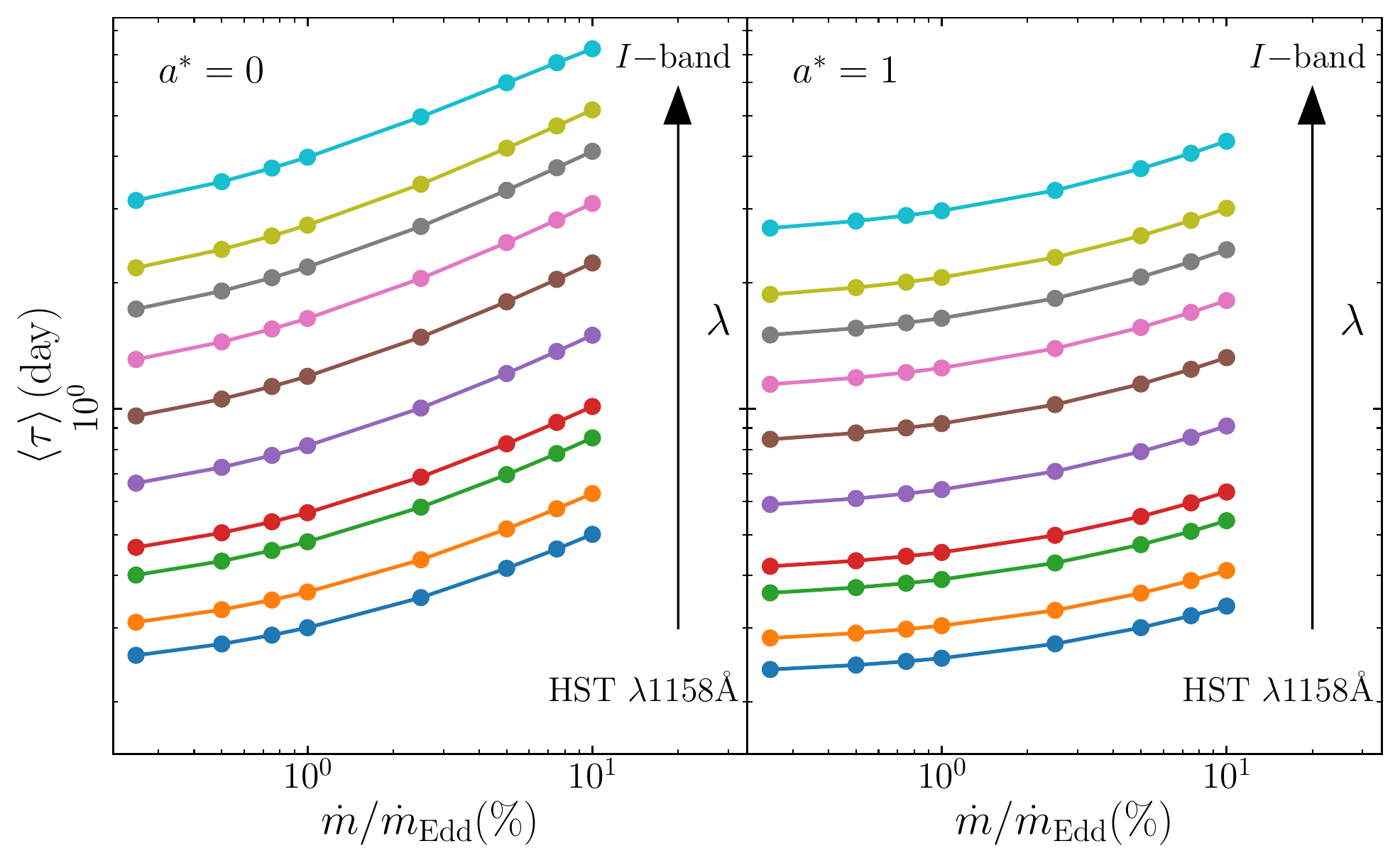}
\includegraphics[width = 0.49\linewidth]{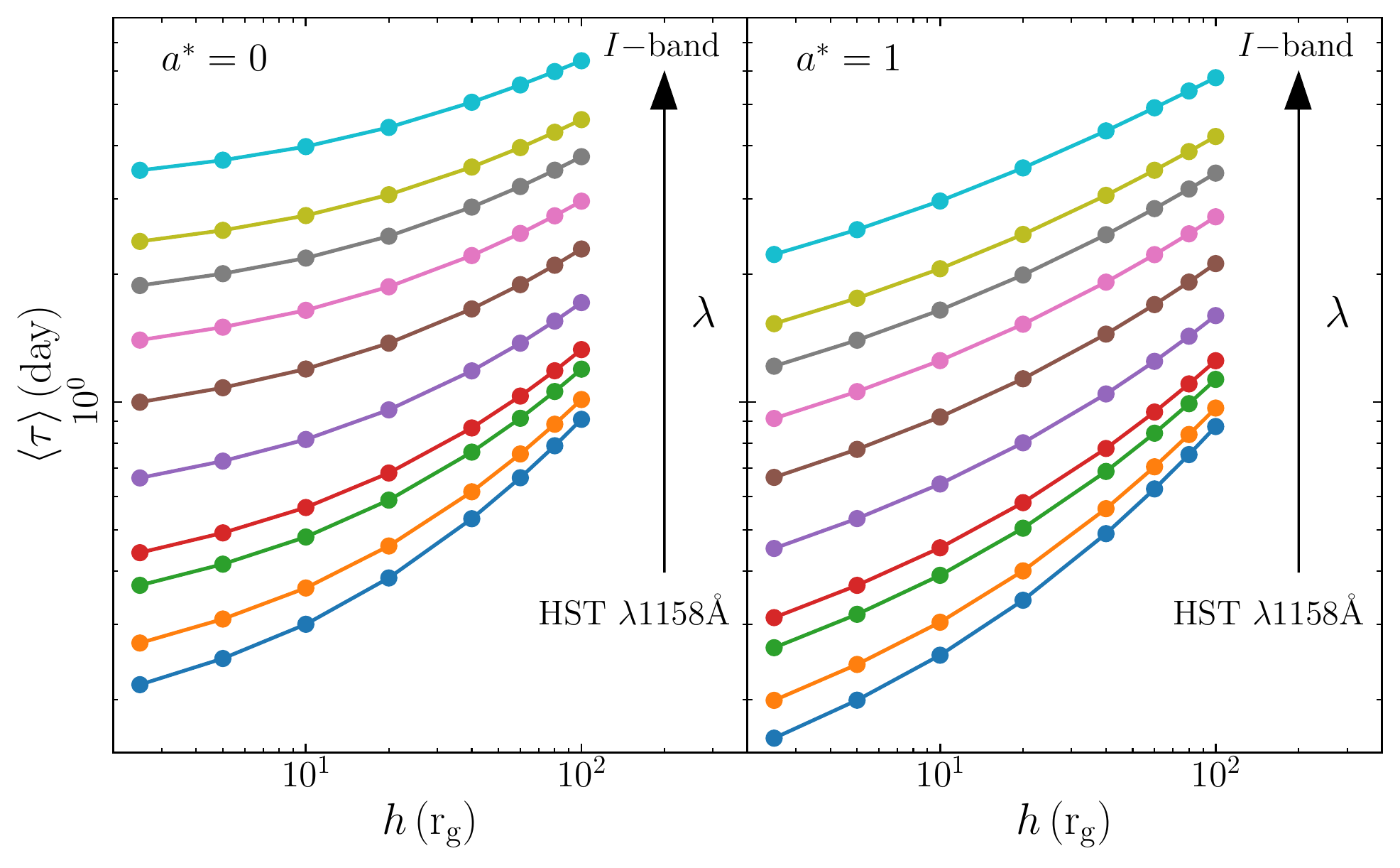}
\caption{Left panel: Mean time-delay as function of \mdot\ ($h=10~\rm r_g$). Right panel: mean time-delay as function of $h$ ($\dot{m}/\dot{m}_{\rm Edd}=1~\%$). 
The left and right plots in each panel correspond to $a^\ast =0$ and 1, respectively. Time delays are shown for all wavebands, increasing $\lambda$ from bottom to top.}
\label{fig:tau_var}
\end{figure*}
%--------

Using these values, we computed $\Psi$, considering eight values for the lamp-post height [$h ({\rm r_g}) =$ 2.5, 5, 10, 20, 40 ,60, 80, 100] and eight values of the accretion rate [$\dot{m}/\dot{m}_{\rm Edd}(\%) =$ 0.25, 0.5, 0.75, 1, 2.5, 5, 7.5, 10]. We assumed a disk density $n_{\rm H} = 10^{17}~\rm cm^{-3}$, and BH spins $a^\ast =0$ and 1. We also consider the following wavebands presented by \cite{Fausnaugh16}: {\it HST} $\lambda1158$, {\it HST} $\lambda1367$, {\it HST} $\lambda1746$, {\it Swift} UVW2, {\it Swift} UVW1, and the \textit{U, B, V, R, I} Johnson-Cousins. We assumed top-hat transmission curves for all filters, with a width of 5~\AA\ for each of the $HST$ bands, 1066~\AA\ and 2892~\AA\ for the {\it R} and {\it I} filters\footnote{\url{https://www.aip.de/en/research/facilities/stella/instruments/data/johnson-ubvri-filter-curves}}, and the widths listed in \citep{Edelson15} for the {\it UVW2, UVW1, U, B} and {\it V} filters.

\subsection{Effects of the accretion rate}
\label{sec:mdotvar}

The left panel of Fig.~\ref{fig:responses} shows the response functions for all the values of \mdot, for $a^\ast = 0$ and 1, in the {\it HST }$\lambda1158$ and the \textit{I} bands (the shortest and longest wavelengths, respectively), for $h=10~\rm r_g$. First, the responses for all accretion rates and in both bands (actually in all bands) start rising at the same time. This is due to the fact that all disk elements emit a BB spectrum and, at the beginning, we observe elements close to the BH whose temperature is such that $\lambda_{\rm max}$ is shorter than 1158~\AA. Consequently, the flux even at the shortest wavelengths will start increasing at the same time.

Second, $\Psi$ increases in amplitude and gets narrower as the accretion rate decreases. The former effect is due to the fact that the disk temperature decreases with decreasing \mdot, thus $F_{\rm NT}(r)$ is smaller. Consequently, for a constant X-ray luminosity,$F_{\rm abs}(r)$ will increase, hence the disk excess flux (i.e., the disk response) will increase for lower \mdot\ values. Regarding the second effect, we note that, in general, $\Psi$ starts decreasing when the temperature of the disk elements, contributing to the observed flux in a given wavelength range, is so low such that the flux comes from the Wien part of the spectrum. As time passes, we observe emission from disk elements which are located further out and are colder. As a result their emission at short wavelengths is diminished, hence the smaller width of response functions at short wavelengths (for all \mdot\ and $h$, as seen in the next section). At the same time, as \mdot\ decreases, the overall temperature (eq.~\ref{eq2}) decreases. Thus, the response function (at a given wavelength) will start decreasing at earlier times, causing the full response function to be narrower for lower \mdot.

%------------------------------------------------------------------------------------
%%%%%%%%%
\subsection{Effects of the lamp-post height}
\label{sec:hvar}

The right panel of Fig.~\ref{fig:responses} shows the response functions for all the values of $h$, for spins 0 and 1, $\dot{m}/\dot{m}_{\rm Edd}=1\%$ and the {\it HST }$\lambda1158$ and the \textit{I} bands. As expected, the larger the height of the lamp-post the later the response function starts, and the longer it lasts. This is due to the light travel time from the X-ray source to the disk. In addition, the amplitude of the response functions increases with height. In fact, the incident flux is proportional to the cosine of the incident angle (defined as the angle between the normal to the disk and the photon trajectory; $\cos \theta_{\rm inc} = h/\sqrt{h^2 + r^2}$). Hence, by increasing the height $\cos \theta_{\rm inc}$ increases leading to a larger incident flux. This effect is mainly important for large radii, where the disk is completely neutral. At smaller radii, as the height increases, $\cos \theta_{\rm inc}$ still increases but the incident flux decreases due to the increase of the distance of the X-ray source to the disk. At the same time though, the ionization state of the inner parts of the disk decreases, which results in lower $F_{\rm ref}$, hence larger $F_{\rm abs}$. Combining both effects the amplitude of the response function increases with height.

We note that the response functions at low spin are broader and have lower amplitudes than the ones for $a^\ast =1$. This is due to the fact that for the same \mdot/\mdot$_{\rm Edd}$ value, the physical value of \mdot\ (in $\rm M_{\odot}~yr^{-1}$) is higher in the low spin case\footnote{Since $\dot{m} = L/\eta c^2$, and the radiative efficiency $\eta$ is smaller for a low spin, \mdot\ (in physical units) is larger than for $a^\ast = 1$.}. Consequently the disk is hotter and $\Psi$ will have a lower amplitude and will be broader, as explained in the previous Section. 

\section{The time delays}
\label{sec:tau}

Then, we estimate the centroid time delay of the transfer functions at a give wavelength ($\lambda$) as follows,
\begin{equation}
\langle \tau(\lambda) \rangle = \frac{\int \tau \Psi(\tau, \lambda) \rm d\tau}{\int  \Psi(\tau, \lambda) \rm d\tau}.
\end{equation}
\noindent
It is this mean time lag that we can compare with the observed time lags between X-rays and UV/optical light curves.The left panels of Fig.~\ref{fig:tau_var} show the dependence of \meantau\ on \mdot\ (for $a^\ast = 0$ and 1; $h=10~\rm r_g$). The curves in the leftmost panel show that, at a given wavelength, the mean time lag increases with increasing \mdot. This is due to the fact that the width of the response increases with increasing \mdot. The mean time lag increases with the same rate in all wavebands. In addition, \meantau\ is larger and increases in a steeper way with increasing \mdot\ for a non-rotating BH compared to a maximally rotating one. This is due to the fact that the response width is larger and increases more with increasing \mdot\ for $a^\ast = 0$ (see Fig.~\ref{fig:responses}).

The right panels of Fig.~\ref{fig:tau_var} shows the mean time lags as a function of source height ($\dot{m}/\dot{m}_{\rm Edd}=1\%$). The mean time lag increases with increase source height, as expected (since the respective responses are delayed and last longer as the height increases; see the right panels in Fig.~\ref{fig:responses}). The time lags at a given energy band are slightly larger in the $a^\ast=0$ case, because the respective responses are wider.

\section{The case of NGC 5548}
\label{sec:fit}

\begin{figure}
\centering
\includegraphics[width = 0.99\linewidth]{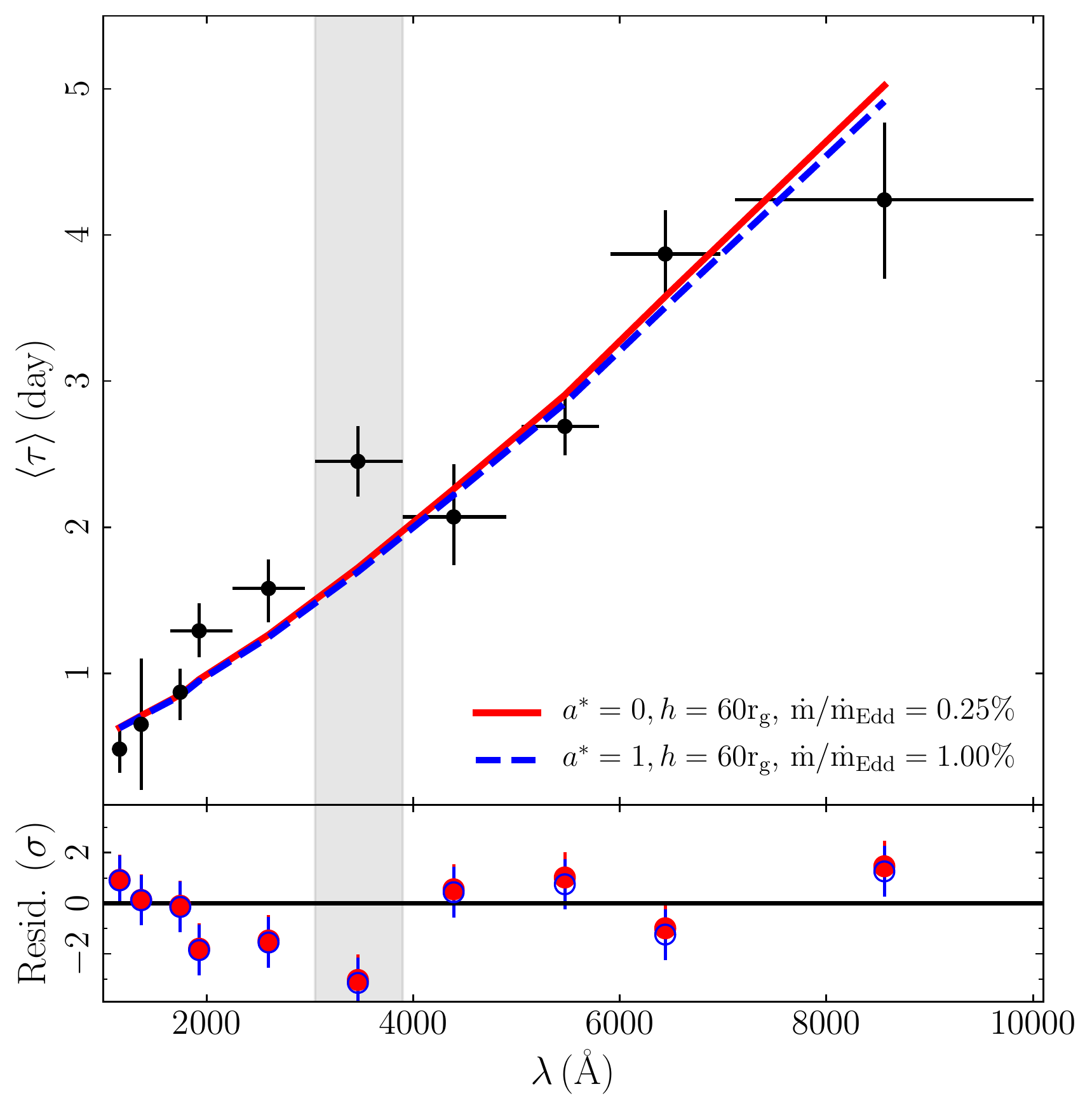}
\caption{Best-fit models for $a^\ast = 0$ (solid line) and $a^\ast = 1$ (dashed line) obtained by fitting the observed time lags, excluding the measurement in the $U-$band (shaded region; see Section~\ref{sec:fit} for details).}
\label{fig:tau_lambda}
\end{figure}

The filled points in Fig.~\ref{fig:tau_lambda} represent the observed time lags between the X-ray and the UV/optical light curves in NGC~5548. They have been estimated by adding 0.65 day (i.e., the observed time lag between X--rays and the HST$\lambda 1367$ light curve) to the time lags listed in Table 6 of \cite{Fausnaugh16}. We compared the observed time lags to our model predictions as follows. 

We estimated the $\chi^2$ between the model lag-spectra (for all \mdot\ and heights; 64 in total for each spin) and the data, and we chose the model time-lags with the minimum $\chi^2$ value ($\chi^2_{\rm min}$). The fit was not statistically acceptable neither for $a^\ast =0$ nor for $a^\ast =1$, $\chi^2_{\rm min}=19.9/8$ degrees of freedom (dof), and 19.2/8, respectively. This is most likely due to the fact that the time lag in the $U-$band appears to be larger compared to the general trend. This was already noticed by \cite{Edelson15} and \cite{Fausnaugh16}, and is probably due to an additional delay caused by the BLR \citep[known as the `Balmer jump'][]{Korista01}. We re-fitted the observed lags by excluding the $U-$band point from the fit. In this case, we obtained statistically accepted fits with $\rm chi^2_{\rm min}/dof=10.8/7$ and $10.7/7$ ($p-$value=0.15 and 0.16) for $a^\ast=0$ and $a^\ast=1$, respectively. The solid (dashed) lines in the same figure indicate the best-fit model to the observed lag spectrum for $a^\ast=0$ ($a^\ast=1$). We note that some deficit can be seen in the $I-$band, though not statistically significant. This might be due to $r_{\rm out}$ being smaller than $10^4~\rm r_g$.

%--------Fig 3
%--------

The best-fit values of height and accretion rate are (${\rm 60~r_g, 0.25\%}$) and ($\rm 60~r_g, 1\%$) for $a^\ast =0$ and $a^\ast =1$, respectively. The best-fit \mdot\ in the case of a non-rotating BH coincides with the lowest value we considered, with a 3-$\sigma$ upper limit of 2.5\%. The 1-$\sigma$ confidence region of \mdot\ for $a^\ast=1$ is $0.25-2.5\%$. The best-fit accretion rates correspond to 0.0048 and $0.0026 ~\rm M_\odot/year$ for $a^\ast =0$ and $a^\ast =1$, respectively. Their difference is smaller than the difference of the best fit values in Eddington units. The best-fit heights are identical in both cases, with the 3-$\sigma$ confidence region being ($40 - 80~\rm r_g$) and ($20-80~\rm r_g$) for $a^\ast =0$ and $a^\ast =1$, respectively.

\section{Conclusions}

We calculate the disk response functions in various wavebands when it is illuminated by X-rays assuming a lamp-post geometry. We considere all relativistic effects in the light propagation from the X-ray source to the disk and from the disk to the observer. We also account for the disk X-ray reflection by computing the disk ionization at each radius. We found that: i) the delays between X-rays and optical/UV bands increase with increasing source height and increasing accretion rate, and ii) the disk response in all UV/optical bands increases when the source height increases and the accretion rate decreases. Therefore, we do not expect a strong thermal reverberation signal in objects with high accretion rate and strong X-ray reflection signatures like, for example, the X-ray bright narrow-line Seyfert-1 galaxies. 

Using reasonable values for the model parameters (i.e., BH mass, inclination, X-ray spectral slope and mean flux) we explained the observed time-lag spectrum in NGC 5548. The best-fit results indicate a source height larger than 20 or $40~ \rm r_g$ (3-$\sigma$ limit for $a^\ast=1$ or 0, respectively). This is significantly larger than $4-5~\rm r_g$ which is the height estimate from the modeling of the X-ray time-lags in a few bright Seyferts, assuming the same geometry \citep[i.e.,][]{Emmanoulopoulos2014, Epitropakis16, Chainakun16, Caballero18}. Nevertheless, our results are in agreement with \cite{Brenneman12} who inferred a height of the X-ray source in NGC 5548 $\sim 100~ \rm r_g$. An alternative solution is provided by \cite{Gardner17}. Their model consists of a puffed-up, Comptonized inner disk region. They proposed that that the continuum UV/optical lags are due to the expansion/contraction of this region, in both radius and height, in response to the variable X-ray heating of its inner edge. We also note that our results depend on the assumption of a thin, plane-parallel Novikov-Thorne disk. Different disk geometries \citep[a tilted disk for example;][]{Nealon15} might affect the source height estimation, however exploring this goes beyond the scope of our work.

As for the accretion rate, the best-fit values indicate rates which are $\sim 1$ per cent of the Eddington limit. According to \cite{Fausnaugh16}, the mean source flux at 5100~\AA\ is $\sim 4.6\times 10^{-11}~\rm erg ~ s^{-1}~cm^{-2}$. Assuming a bolometric correction factor of $7.8\pm1.7$ \citep{Krawczyk13}, this implies a bolometric luminosity of $(2.5 \pm 0.5) \times 10^{44} \rm ~erg ~ s^{-1}$, which is $0.034\pm0.008$ of the Eddington luminosity limit for $M_{\rm BH} =5.7\times 10^7~\rm M_{\odot}$. This value is at odds with the \mdot\ estimate for $a^\ast = 0$ (the $3\sigma$ upper limit is just 0.025) but entirely consistent with the \mdot\ estimate considering a standard accretion disk around a maximally rotating BH.

Contrary to our results, \cite{Starkey17} found that a standard disk, with a low \mdot, is ruled out by the data. A significant difference between theirs and our modeling is that they kept $h$ fixed at $6~\rm r_g$ and let the inclination free, while we fixed the inclination to 40\degr\ and let $h$ free. In addition, we do not assume a fixed albedo for the disk, but we compute the flux that is reflected by the disk, at each radius, based on its ionization state accounting for all relativistic effects. Furthermore, we compute the time lags using the model disk response functions, without assuming that all variations in the UV/optical light curves are due to thermal reverberation. In the future, we plan to compare the model with the observed UV/optical light curves, as this will be a crucial test for the model.

%In the future, we plan to investigate how the dependence of the disk response functions on $M_{\rm BH}$, inclination, X-ray spectral shape, disk density, inner and outer radii of the disk, and $L_{\rm X}/L_{\rm Edd}$. We note that our preliminary simulations show that the dependence of $\Psi$ on $L_{\rm X}/L_{\rm Edd}$ is not linear, however more in-depth investigations are needed. We also plan to construct model UV/optical light curves and compare them with the observed ones in NGC 5548 and in other sources with well sampled light curves, like NGC 4593 \citep{Mchardy18, Cackett18}, Mrk~509 and NGC~4151 \cite{Edelson19}.

%%%%%%%%%%%%%%%%%%%%%%%%%%%%%%%%%%%%%%%%%%%%
\begin{acknowledgements}

MD thanks MEYS of Czech Republic for the support through the 18-00533S research project and his home institution, ASU, supported by the project RVO:67985815. IP would like to thank ASU for hospitality.
\end{acknowledgements}

% WARNING
%-------------------------------------------------------------------
%	\bibliographystyle{aasjournal}
%	\bibliography{references}

\begin{thebibliography}{}
\expandafter\ifx\csname natexlab\endcsname\relax\def\natexlab#1{#1}\fi
\providecommand{\url}[1]{\href{#1}{#1}}
\providecommand{\dodoi}[1]{doi:~\href{http://doi.org/#1}{\nolinkurl{#1}}}
\providecommand{\doeprint}[1]{\href{http://ascl.net/#1}{\nolinkurl{http://ascl.net/#1}}}
\providecommand{\doarXiv}[1]{\href{https://arxiv.org/abs/#1}{\nolinkurl{https://arxiv.org/abs/#1}}}

\bibitem[{{Bentz} \& {Katz}(2015)}]{Bentz15}
{Bentz}, M.~C., \& {Katz}, S. 2015, \pasp, 127, 67, \dodoi{10.1086/679601}

\bibitem[{{Brenneman} {et~al.}(2012){Brenneman}, {Elvis}, {Krongold}, {Liu}, \&
  {Mathur}}]{Brenneman12}
{Brenneman}, L.~W., {Elvis}, M., {Krongold}, Y., {Liu}, Y., \& {Mathur}, S.
  2012, \apj, 744, 13, \dodoi{10.1088/0004-637X/744/1/13}

\bibitem[{{Caballero-Garc{\'{\i}}a} {et~al.}(2018){Caballero-Garc{\'{\i}}a},
  {Papadakis}, {Dov{\v c}iak}, {Bursa}, {Epitropakis}, {Karas}, \&
  {Svoboda}}]{Caballero18}
{Caballero-Garc{\'{\i}}a}, M.~D., {Papadakis}, I.~E., {Dov{\v c}iak}, M.,
  {et~al.} 2018, \mnras, 480, 2650, \dodoi{10.1093/mnras/sty1990}

\bibitem[{{Cackett} {et~al.}(2018){Cackett}, {Chiang}, {McHardy}, {Edelson},
  {Goad}, {Horne}, \& {Korista}}]{Cackett18}
{Cackett}, E.~M., {Chiang}, C.-Y., {McHardy}, I., {et~al.} 2018, \apj, 857, 53,
  \dodoi{10.3847/1538-4357/aab4f7}

\bibitem[{{Chainakun} {et~al.}(2016){Chainakun}, {Young}, \&
  {Kara}}]{Chainakun16}
{Chainakun}, P., {Young}, A.~J., \& {Kara}, E. 2016, \mnras, 460, 3076,
  \dodoi{10.1093/mnras/stw1105}

\bibitem[{{Chartas} {et~al.}(2009){Chartas}, {Kochanek}, {Dai}, {Poindexter},
  \& {Garmire}}]{Chartas09}
{Chartas}, G., {Kochanek}, C.~S., {Dai}, X., {Poindexter}, S., \& {Garmire}, G.
  2009, \apj, 693, 174, \dodoi{10.1088/0004-637X/693/1/174}

\bibitem[{{De Marco} {et~al.}(2013){De Marco}, Ponti, Cappi, Dadina, Uttley,
  Cackett, Fabian, \& Miniutti}]{DeMarco2013}
{De Marco}, B., Ponti, G., Cappi, M., {et~al.} 2013, \mnras, 431, 2441,
  \dodoi{10.1093/mnras/stt339}

\bibitem[{{Edelson} {et~al.}(2015){Edelson}, {Gelbord}, {Horne}, {McHardy},
  {Peterson}, {Ar{\'e}valo}, {Breeveld}, {De Rosa}, {Evans}, {Goad}, {Kriss},
  {Brandt}, {Gehrels}, {Grupe}, {Kennea}, {Kochanek}, {Nousek}, {Papadakis},
  {Siegel}, {Starkey}, {Uttley}, {Vaughan}, {Young}, {Barth}, {Bentz},
  {Brewer}, {Crenshaw}, {Dalla Bont{\`a}}, {De Lorenzo-C{\'a}ceres}, {Denney},
  {Dietrich}, {Ely}, {Fausnaugh}, {Grier}, {Hall}, {Kaastra}, {Kelly},
  {Korista}, {Lira}, {Mathur}, {Netzer}, {Pancoast}, {Pei}, {Pogge},
  {Schimoia}, {Treu}, {Vestergaard}, {Villforth}, {Yan}, \& {Zu}}]{Edelson15}
{Edelson}, R., {Gelbord}, J.~M., {Horne}, K., {et~al.} 2015, \apj, 806, 129,
  \dodoi{10.1088/0004-637X/806/1/129}

\bibitem[{{Edelson} {et~al.}(2019){Edelson}, {Gelbord}, {Cackett}, {Peterson},
  {Horne}, {Barth}, {Starkey}, {Bentz}, {Brandt}, {Goad}, {Joner}, {Korista},
  {Netzer}, {Page}, {Uttley}, {Vaughan}, {Breeveld}, {Cenko}, {Done}, {Evans},
  {Fausnaugh}, {Ferland}, {Gonzalez-Buitrago}, {Gropp}, {Grupe}, {Kaastra},
  {Kennea}, {Kriss}, {Mathur}, {Mehdipour}, {Mudd}, {Nousek}, {Schmidt},
  {Vestergaard}, \& {Villforth}}]{Edelson19}
{Edelson}, R., {Gelbord}, J., {Cackett}, E., {et~al.} 2019, \apj, 870, 123,
  \dodoi{10.3847/1538-4357/aaf3b4}

\bibitem[{Emmanoulopoulos {et~al.}(2014)Emmanoulopoulos, Papadakis,
  Dov{\v{c}}iak, \& McHardy}]{Emmanoulopoulos2014}
Emmanoulopoulos, D., Papadakis, I.~E., Dov{\v{c}}iak, M., \& McHardy, I.~M.
  2014, \mnras, 439, 22, \dodoi{10.1093/mnras/stu249}

\bibitem[{{Epitropakis} {et~al.}(2016){Epitropakis}, {Papadakis}, {Dov{\v
  c}iak}, {Pech{\'a}{\v c}ek}, {Emmanoulopoulos}, {Karas}, \&
  {McHardy}}]{Epitropakis16}
{Epitropakis}, A., {Papadakis}, I.~E., {Dov{\v c}iak}, M., {et~al.} 2016, \aap,
  594, A71, \dodoi{10.1051/0004-6361/201527748}

\bibitem[{{Fabian} {et~al.}(2009){Fabian}, {Zoghbi}, {Ross}, {Uttley}, {Gallo},
  {Brandt}, {Blustin}, {Boller}, {Caballero-Garcia}, {Larsson}, {Miller},
  {Miniutti}, {Ponti}, {Reis}, {Reynolds}, {Tanaka}, \& {Young}}]{Fabian2009}
{Fabian}, A.~C., {Zoghbi}, A., {Ross}, R.~R., {et~al.} 2009, \nat, 459, 540,
  \dodoi{10.1038/nature08007}

\bibitem[{{Fausnaugh} {et~al.}(2016){Fausnaugh}, {Denney}, {Barth}, {Bentz},
  {Bottorff}, {Carini}, {Croxall}, {De Rosa}, {Goad}, {Horne}, {Joner},
  {Kaspi}, {Kim}, {Klimanov}, {Kochanek}, {Leonard}, {Netzer}, {Peterson},
  {Schn{\"u}lle}, {Sergeev}, {Vestergaard}, {Zheng}, {Zu}, {Anderson},
  {Ar{\'e}valo}, {Bazhaw}, {Borman}, {Boroson}, {Brandt}, {Breeveld}, {Brewer},
  {Cackett}, {Crenshaw}, {Dalla Bont{\`a}}, {De Lorenzo-C{\'a}ceres},
  {Dietrich}, {Edelson}, {Efimova}, {Ely}, {Evans}, {Filippenko}, {Flatland},
  {Gehrels}, {Geier}, {Gelbord}, {Gonzalez}, {Gorjian}, {Grier}, {Grupe},
  {Hall}, {Hicks}, {Horenstein}, {Hutchison}, {Im}, {Jensen}, {Jones},
  {Kaastra}, {Kelly}, {Kennea}, {Kim}, {Korista}, {Kriss}, {Lee}, {Lira},
  {MacInnis}, {Manne-Nicholas}, {Mathur}, {McHardy}, {Montouri}, {Musso},
  {Nazarov}, {Norris}, {Nousek}, {Okhmat}, {Pancoast}, {Papadakis}, {Parks},
  {Pei}, {Pogge}, {Pott}, {Rafter}, {Rix}, {Saylor}, {Schimoia}, {Siegel},
  {Spencer}, {Starkey}, {Sung}, {Teems}, {Treu}, {Turner}, {Uttley},
  {Villforth}, {Weiss}, {Woo}, {Yan}, \& {Young}}]{Fausnaugh16}
{Fausnaugh}, M.~M., {Denney}, K.~D., {Barth}, A.~J., {et~al.} 2016, \apj, 821,
  56, \dodoi{10.3847/0004-637X/821/1/56}

\bibitem[{{Garc{\'{\i}}a} {et~al.}(2016){Garc{\'{\i}}a}, {Fabian}, {Kallman},
  {Dauser}, {Parker}, {McClintock}, {Steiner}, \& {Wilms}}]{Garcia16}
{Garc{\'{\i}}a}, J.~A., {Fabian}, A.~C., {Kallman}, T.~R., {et~al.} 2016,
  \mnras, 462, 751, \dodoi{10.1093/mnras/stw1696}

\bibitem[{Gardner \& Done(2017)}]{Gardner17}
Gardner, E., \& Done, C. 2017, Monthly Notices of the Royal Astronomical
  Society, 470, 3591, \dodoi{10.1093/mnras/stx946}

\bibitem[{Haardt(1993)}]{Haardt93}
Haardt, F. 1993, \apj, 413, 680, \dodoi{10.1086/173036}

\bibitem[{{Kammoun} {et~al.}(2019){Kammoun}, {Dom{\v c}ek}, {Svoboda}, {Dov{\v
  c}iak}, \& {Matt}}]{Kammoun19}
{Kammoun}, E.~S., {Dom{\v c}ek}, V., {Svoboda}, J., {Dov{\v c}iak}, M., \&
  {Matt}, G. 2019, \mnras, 485, 239, \dodoi{10.1093/mnras/stz408}

\bibitem[{{Korista} \& {Goad}(2001)}]{Korista01}
{Korista}, K.~T., \& {Goad}, M.~R. 2001, \apj, 553, 695, \dodoi{10.1086/320964}

\bibitem[{{Krawczyk} {et~al.}(2013){Krawczyk}, {Richards}, {Mehta}, {Vogeley},
  {Gallagher}, {Leighly}, {Ross}, \& {Schneider}}]{Krawczyk13}
{Krawczyk}, C.~M., {Richards}, G.~T., {Mehta}, S.~S., {et~al.} 2013, \apjs,
  206, 4, \dodoi{10.1088/0067-0049/206/1/4}

\bibitem[{{Mathur} {et~al.}(2017){Mathur}, {Gupta}, {Page}, {Pogge},
  {Krongold}, {Goad}, {Adams}, {Anderson}, {Ar{\'e}valo}, {Barth}, {Bazhaw},
  {Beatty}, {Bentz}, {Bigley}, {Bisogni}, {Borman}, {Boroson}, {Bottorff},
  {Brandt}, {Breeveld}, {Brown}, {Brown}, {Cackett}, {Canalizo}, {Carini},
  {Clubb}, {Comerford}, {Coker}, {Corsini}, {Crenshaw}, {Croft}, {Croxall},
  {Dalla Bont{\`a}}, {Deason}, {Denney}, {De Lorenzo-C{\'a}ceres}, {De Rosa},
  {Dietrich}, {Edelson}, {Ely}, {Eracleous}, {Evans}, {Fausnaugh}, {Ferland},
  {Filippenko}, {Flatland}, {Fox}, {Gates}, {Gehrels}, {Geier}, {Gelbord},
  {Gorjian}, {Greene}, {Grier}, {Grupe}, {Hall}, {Henderson}, {Hicks},
  {Holmbeck}, {Holoien}, {Horenstein}, {Horne}, {Hutchison}, {Im}, {Jensen},
  {Johnson}, {Joner}, {Jones}, {Kaastra}, {Kaspi}, {Kelly}, {Kelly}, {Kennea},
  {Kim}, {Kim}, {Kim}, {King}, {Klimanov}, {Kochanek}, {Korista}, {Kriss},
  {Lau}, {Lee}, {Leonard}, {Li}, {Lira}, {Ma}, {MacInnis}, {Manne-Nicholas},
  {Malkan}, {Mauerhan}, {McGurk}, {McHardy}, {Montouri}, {Morelli}, {Mosquera},
  {Mudd}, {Muller-Sanchez}, {Musso}, {Nazarov}, {Netzer}, {Nguyen}, {Norris},
  {Nousek}, {Ochner}, {Okhmat}, {Ou-Yang}, {Pancoast}, {Papadakis}, {Parks},
  {Pei}, {Peterson}, {Pizzella}, {Poleski}, {Pott}, {Rafter}, {Rix}, {Runnoe},
  {Saylor}, {Schimoia}, {Schn{\"u}lle}, {Sergeev}, {Shappee}, {Shivvers},
  {Siegel}, {Simonian}, {Siviero}, {Skielboe}, {Somers}, {Spencer}, {Starkey},
  {Stevens}, {Sung}, {Tayar}, {Tejos}, {Turner}, {Uttley}, {Van Saders},
  {Vestergaard}, {Vican}, {Villanueva}, {Villforth}, {Weiss}, {Woo}, {Yan},
  {Young}, {Yuk}, {Zheng}, {Zhu}, \& {Zu}}]{Mathur17}
{Mathur}, S., {Gupta}, A., {Page}, K., {et~al.} 2017, \apj, 846, 55,
  \dodoi{10.3847/1538-4357/aa832b}

\bibitem[{{McHardy} {et~al.}(2014){McHardy}, {Cameron}, {Dwelly}, {Connolly},
  {Lira}, {Emmanoulopoulos}, {Gelbord}, {Breedt}, {Arevalo}, \&
  {Uttley}}]{Mchardy14}
{McHardy}, I.~M., {Cameron}, D.~T., {Dwelly}, T., {et~al.} 2014, \mnras, 444,
  1469, \dodoi{10.1093/mnras/stu1636}

\bibitem[{{McHardy} {et~al.}(2018){McHardy}, {Connolly}, {Horne}, {Cackett},
  {Gelbord}, {Peterson}, {Pahari}, {Gehrels}, {Goad}, {Lira}, {Arevalo},
  {Baldi}, {Brandt}, {Breedt}, {Chand}, {Dewangan}, {Done}, {Elvis},
  {Emmanoulopoulos}, {Fausnaugh}, {Kaspi}, {Kochanek}, {Korista}, {Papadakis},
  {Rao}, {Uttley}, {Vestergaard}, \& {Ward}}]{Mchardy18}
{McHardy}, I.~M., {Connolly}, S.~D., {Horne}, K., {et~al.} 2018, \mnras, 480,
  2881, \dodoi{10.1093/mnras/sty1983}

\bibitem[{{Nealon} {et~al.}(2015){Nealon}, {Price}, \& {Nixon}}]{Nealon15}
{Nealon}, R., {Price}, D.~J., \& {Nixon}, C.~J. 2015, \mnras, 448, 1526,
  \dodoi{10.1093/mnras/stv014}

\bibitem[{{Novikov} \& {Thorne}(1973)}]{Novikov73}
{Novikov}, I.~D., \& {Thorne}, K.~S. 1973, in Black Holes (Les Astres Occlus),
  ed. C.~{Dewitt} \& B.~S. {Dewitt}, 343--450

\bibitem[{{Reis} \& {Miller}(2013)}]{Reis13}
{Reis}, R.~C., \& {Miller}, J.~M. 2013, \apjl, 769, L7,
  \dodoi{10.1088/2041-8205/769/1/L7}

\bibitem[{{Shakura} \& {Sunyaev}(1973)}]{Shakura73}
{Shakura}, N.~I., \& {Sunyaev}, R.~A. 1973, \aap, 24, 337

\bibitem[{{Shapiro} {et~al.}(1976){Shapiro}, {Lightman}, \&
  {Eardley}}]{Shapiro76}
{Shapiro}, S.~L., {Lightman}, A.~P., \& {Eardley}, D.~M. 1976, \apj, 204, 187,
  \dodoi{10.1086/154162}

\bibitem[{{Shappee} {et~al.}(2014){Shappee}, {Prieto}, {Grupe}, {Kochanek},
  {Stanek}, {De Rosa}, {Mathur}, {Zu}, {Peterson}, {Pogge}, {Komossa}, {Im},
  {Jencson}, {Holoien}, {Basu}, {Beacom}, {Szczygie{\l}}, {Brimacombe},
  {Adams}, {Campillay}, {Choi}, {Contreras}, {Dietrich}, {Dubberley},
  {Elphick}, {Foale}, {Giustini}, {Gonzalez}, {Hawkins}, {Howell}, {Hsiao},
  {Koss}, {Leighly}, {Morrell}, {Mudd}, {Mullins}, {Nugent}, {Parrent},
  {Phillips}, {Pojmanski}, {Rosing}, {Ross}, {Sand}, {Terndrup}, {Valenti},
  {Walker}, \& {Yoon}}]{Shappee14}
{Shappee}, B.~J., {Prieto}, J.~L., {Grupe}, D., {et~al.} 2014, \apj, 788, 48,
  \dodoi{10.1088/0004-637X/788/1/48}

\bibitem[{{Starkey} {et~al.}(2017){Starkey}, {Horne}, {Fausnaugh}, {Peterson},
  {Bentz}, {Kochanek}, {Denney}, {Edelson}, {Goad}, {De Rosa}, {Anderson},
  {Ar{\'e}valo}, {Barth}, {Bazhaw}, {Borman}, {Boroson}, {Bottorff}, {Brandt},
  {Breeveld}, {Cackett}, {Carini}, {Croxall}, {Crenshaw}, {Dalla Bont{\`a}},
  {De Lorenzo-C{\'a}ceres}, {Dietrich}, {Efimova}, {Ely}, {Evans},
  {Filippenko}, {Flatland}, {Gehrels}, {Geier}, {Gelbord}, {Gonzalez},
  {Gorjian}, {Grier}, {Grupe}, {Hall}, {Hicks}, {Horenstein}, {Hutchison},
  {Im}, {Jensen}, {Joner}, {Jones}, {Kaastra}, {Kaspi}, {Kelly}, {Kennea},
  {Kim}, {Kim}, {Klimanov}, {Korista}, {Kriss}, {Lee}, {Leonard}, {Lira},
  {MacInnis}, {Manne-Nicholas}, {Mathur}, {McHardy}, {Montouri}, {Musso},
  {Nazarov}, {Norris}, {Nousek}, {Okhmat}, {Pancoast}, {Parks}, {Pei}, {Pogge},
  {Pott}, {Rafter}, {Rix}, {Saylor}, {Schimoia}, {Schn{\"u}lle}, {Sergeev},
  {Siegel}, {Spencer}, {Sung}, {Teems}, {Turner}, {Uttley}, {Vestergaard},
  {Villforth}, {Weiss}, {Woo}, {Yan}, {Young}, {Zheng}, \& {Zu}}]{Starkey17}
{Starkey}, D., {Horne}, K., {Fausnaugh}, M.~M., {et~al.} 2017, \apj, 835, 65,
  \dodoi{10.3847/1538-4357/835/1/65}

\bibitem[{Svoboda {et~al.}(2012)Svoboda, Dov{\v{c}}iak, Goosmann, Jethwa,
  Karas, Miniutti, \& Guainazzi}]{Svoboda12}
Svoboda, J., Dov{\v{c}}iak, M., Goosmann, R.~W., {et~al.} 2012, Astronomy {\&}
  Astrophysics, 545, A106, \dodoi{10.1051/0004-6361/201219701}

\bibitem[{{Wenger} {et~al.}(2000){Wenger}, {Ochsenbein}, {Egret}, {Dubois},
  {Bonnarel}, {Borde}, {Genova}, {Jasniewicz}, {Lalo{\"e}}, {Lesteven}, \&
  {Monier}}]{Simbad}
{Wenger}, M., {Ochsenbein}, F., {Egret}, D., {et~al.} 2000, \aaps, 143, 9,
  \dodoi{10.1051/aas:2000332}

\end{thebibliography}
%-------------------------------------------------------------------

\end{document}